\documentclass[conference]{IEEEtran}

\usepackage{graphicx}                   
\usepackage{times,amsmath,amssymb,amsfonts}   
\usepackage{bm}                         
\usepackage{psfrag}                     
\usepackage{multirow}
\usepackage{cite}
\usepackage[capitalize]{cleveref}
\usepackage[table]{xcolor}
\usepackage{gensymb}

\usepackage{fancyhdr,lipsum}

\fancypagestyle{mahmood}{%
   \fancyhf{} 
   
   \fancyhead[C]{W. Ahmad, G. Alyami, and I. N. Kostanic, \textquotedblleft Multiuser separation and performance analysis of millimeter wave channels with linear precoding,\textquotedblright ~\textit{accepted} in IEEE International Conference on Communication, Networks and Satellite (COMNETSAT) (IEEE COMNETSAT 2016), Surabaya, Indonesia, Dec. 2016, pp. 66-70.}
}%

\makeatletter
\let\ps@IEEEtitlepagestyle\ps@mahmood
\makeatother

\definecolor{tui-green}{rgb}{0,0.455,0.478}
\definecolor{tui-blue}{rgb}{0,0.2,0.349}
\definecolor{tui-orange}{rgb}{1.0,0.475,0}
\definecolor{tui-lightblue}{rgb}{0.706,0.863,0.863}

\usepackage[load-configurations=abbreviations,load-configurations=binary]{siunitx}	
\DeclareSIUnit{\mms}{\milli\squaremetre}
\DeclareSIUnit{\inch}{in}
\DeclareSIUnit{\inchs}{in\squared}
\DeclareSIUnit{\mil}{mil}
\DeclareSIUnit{\Msps}{Msps}
\DeclareSIUnit{\Mbps}{Mbps}
\DeclareSIUnit{\LSB}{LSB}
\DeclareSIUnit{\pFS}{\percent FS}
\DeclareSIUnit{\dBc}{\deci\bel c}
\DeclareSIUnit{\dBm}{\deci\bel m}
\DeclareSIUnit{\dBFS}{\deci\bel FS}
\DeclareSIUnit{\dB}{\deci\bel}
\DeclareSIUnit{\dBi}{\deci\bel i}
\DeclareSIUnit{\hex}{0x}
\DeclareSIUnit{\vp}{\volt_{\text{p}}}
\DeclareSIUnit{\vpp}{\volt_{\text{pp}}}
\DeclareSIUnit{\kb}{\kilo\bit}
\DeclareSIUnit{\kB}{\kilo\byte}
\DeclareSIUnit{\MB}{\mega\byte}
\DeclareSIUnit{\GHz}{\giga\hertz}
\DeclareSIUnit{\MHz}{\mega\hertz}

\DeclareSIUnit{\mus}{\micro\second}
\DeclareSIUnit{\ns}{\nano\second}
\DeclareSIUnit{\fs}{\femto\second}

\title{Multiuser separation and performance analysis of millimeter wave channels with linear precoding}
\author{
   \IEEEauthorblockN{Waqas Ahmad\IEEEauthorrefmark{1}, Geamel Alyami\IEEEauthorrefmark{2}, Ivica Kostanic \IEEEauthorrefmark{2}}
\IEEEauthorblockA{\IEEEauthorrefmark{1}Huawei Technologies, Pakistan
          \\\{waqas.ahmad1\}@huawei.com}
          \IEEEauthorblockA{\IEEEauthorrefmark{2}Wireless Center of Excellence (WiCE) - Florida Institute of Technology, USA
          \\\{galyami2007\}@my.fit.edu}}

\begin{document}
\maketitle
\IEEEpeerreviewmaketitle
\begin{abstract}
In the conventional multiuser MIMO systems, user selection and scheduling has previously been used as an effective way to increase the sum rate performance of the system. However, the recent concepts of the massive MIMO systems (at centimeter wavelength frequencies) have shown that with higher spatial resolution of antenna arrays different users in the dense scenarios can be spatially separated. This in turn significantly reduces the signal processing efforts required for multiuser selection algorithms. On the other hand, recent measurements at millimeter wave frequencies show that multipath components only arrive from few angular directions leading to high spatial correlation between the paths and co-located users. This paper focus at the investigation of spatial separation among the users at the millimeter wave frequencies with fully digital linear zero-forcing transmit precoding considering various channel propagation parameters. Our analysis results convincingly give a proof that multiuser selection algorithms are still important for millimeter wave communication systems. Results also show that increased number of antenna elements does not give a major benefit to sum rate improvements as compared to the selection of correct number of users to be selected/scheduled.    
\end{abstract}
\vspace{3mm}
\section{Introduction}
Wireless communications at millimeter wave (mmwave) frequencies is one of the major proposals for fifth generation of mobile telephony. The major benefits coupled with mmwave frequencies is the large unused band (30-300 GHz) where \underline{U}ltra \underline{W}ide \underline{B}and (UWB) channel setups could easily be realized at different center frequencies. Additionally, higher carrier frequencies also allow design of compact antenna arrays with large number of antenna elements which increase the directional resolution of the system. Both of the above mentioned features allow multi-gigabyte data rate communications at wireless links.\\   
Very recently, realistic channel sounding campaign~\cite{FGD15} at 2.4 GHz band focused at the spatial separation of multiple users in a densely populated scenario. Measurements were taken with a circular shaped, dual polarized array with large number of antenna elements at the base station (BS). Results in~\cite{FGD15} concluded that under the considered massive antenna setup at the BS, users in the cell can indeed be separated. This is quite intuitive as the dual antenna polarization and circular array geometry capture the Multi-Path Components (MPCs) from distinct Directions Of Arrivals (DOA) which are spatially uncorrelated. These facts have been verified in~\cite{CNR15} for centimeter wavelength frequencies. At the mmwave frequencies, measurement results in~\cite{Rappaport5G,Ref7RapaportmmCapacity} show that MPCs arrive at the receiver from very few distinct directions known as spatial lobes. Hence, the users located very close to each other will almost see the same channel which may lead to higher inter-user correlation. Therefore, at mmwaves the topic of inter-user channel correlation is more interesting and needs thorough investigations. It is important to understand the impact of high directional and bandwidth resolution at the spatial separation among users. At the mmwave 60 GHz frequency band, authors in~\cite{NHJ15} have discussed the mutual orthogonality between the users as a function of inter user distance and the number of co-located users in an open square scenario. In another work~\cite{AK10}, spatial separation of users has been discussed as a function of users probability to be in Line of Sight (LOS) to the BS. Results have shown that user separation is difficult under the scenarios where the LOS probability is high. \\ 
From system design perspective, it is well known that transmit precoding and effects of different SNR levels also play a vital role in defining the spatial separation among users. From authors point of view, inter user orthogonality and the channel propagation parameters that define the degree of separation among the users yet needs to be extensively analyzed for different transmit precoding schemes and SNR levels. In this paper, we have provided a deep insight of the sum rate analysis by changing the channel propagation parameters e.g, number of multipath clusters, number of BS antennas and BS antennas separation for indoor (shopping mall) scenario.
We have used linear zero forcing (ZF) precoder for the analysis of multi-user system. Although linear digital precoding is not yet realized to be used at mmwave frequencies due to increased hardware complexity in context of required number of RF chains. Therefore, hybrid (analog+digital) precoding techniques have received more attention among the wireless communications research community. These hybrid techniques have shown slight performance losses as compared to fully digital linear procodings. Therefore, fully digital ZF precoding just give us a performance upper bound and help to understand the behavior of the channel in a simplistic manner. Additionally, we have used the model in~\cite{Ref7RapaportmmCapacity} for the number of multipath clusters in the mmwave propagation channel. We have investigated the sum rate improvements if the count of multipath cluster is more than the number defined in model\cite{Ref7RapaportmmCapacity} itself.  

\section{Channel Model}
\label{sec:Sysmod}
\begin{figure}[t!]
 \begin{center}
 \includegraphics[width=0.90\columnwidth]{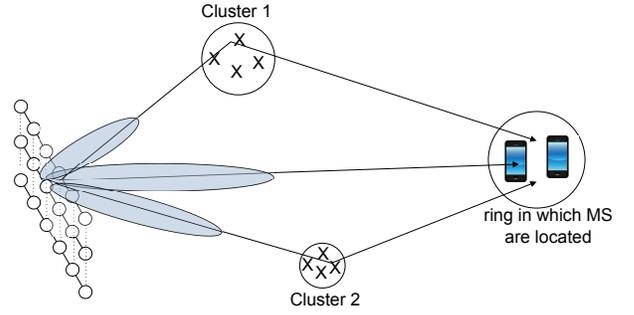}
 \end{center}
 \caption{Channel model and experiment description}
 \label{scenario}
 \end{figure}

Consider a downlink multiuser MIMO system with $n_\text{BS}$ antennas mounted at the base station (BS) of height $h_\text{BS}$ and $n_\text{U}$ be the number of mobile users (MU) each of height $h_\text{U}$. All the users are closely located in the XY plane of the Cartesian coordinate system inside a ring of radius $R$ as shown in Fig.~\ref{scenario}. 
The position of users in the ring follows uniform random distribution. The heights of the BS and MU correspond to z-axis of the coordinate system. Let $ N_{cl} $ be the number of multipath clusters with $ N_{ray} $ subpaths  in each cluster. For simplicity, we assume single bounce multipath clusters. Let $\theta_{i,l}^\text{A} $ and $\theta_{i,l}^\text{D}$ be azimuth angle of arrival and departure of $l^\text{th}$ ray of the $i^\text{th}$ cluster respectively. Similarly $\phi_{i,l}^\text{A} $ and $\phi_{i,l}^\text{D}$ correspond to the elevation angles of arrival and departure respectively. Let $K$ be the number of antennas at each MU, then the channel matrix $ \bm{H}_i(\tau )\in \mathbb{C}^{n_\text{BS}\times K} $ of a single user MIMO link can be written as 
\begin{equation}
\label{HLTI}
\begin{array} {lcl}
\bm{H}_i\left( \tau\right) & = &  \sum_{i=1}^{N_{cl}} \sum_{l=1}^{N_{ray}} \alpha_{i,l} \sqrt{L\left( r_{i,l}\right) } \bm{a}_r\left( \phi_{i,l}^A , \theta_{i,l}^A\right) \times \\ &  &  \bm{a}_t^H\left( \phi_{i,l}^D,\theta_{i,l}^D\right) h\left( \tau -\tau_{i,l}\right)  + \bm{H}_\text{LOS}\left( \tau\right) 
\end{array}
\end{equation}
where,
\begin{itemize}
\item $\bm{H}_\text{LOS}\left( \tau\right)$ is the channel impulse response matrix of the LOS component arriving at delay $\tau$.
\item $\bm{a}_t$ and $\bm{a}_r$ are the transmit and receive antenna array steering vectors.
\item $h\left( \tau -\tau_{i,l}\right)$ is the impulse response at the delay tap $\tau_{i,l}$ of the $l^\text{th}$ ray of the $i^\text{th}$ cluster.
\item $\alpha_{i,l}$ and $L\left( r_{i,l}\right)$ are the complex channel gain and attenuation loss of the $l^\text{th}$ ray of the $i^\text{th}$ cluster respectively.
\end{itemize}
Since each subpath may have different scattering conditions so the subpath phases can be considered as independent identical distribution. We have assumed constant distance between the antenna array and all the scatterers which belong to the same cluster to simplify the model. Let $r_{i}$ be the propagation distance between BS and MU via the central ray of the $i^\text{th}$ cluster. The propagation distance of subpaths associated to the $i^\text{th}$ cluster is geometrically computed as 
\begin{equation}
\label{ril}
\begin{array} {lcl}
r_{i,l} =  r_i+ \\  \sqrt{\left( h_\text{BS}-h_\text{U}+ r_i \sin\theta_{i,l}^D\right) ^2+\left( d-r_i\cos\theta_{i,l}^D\cos\phi_{i,l}^D\right) ^2}
\end{array}\cdot
\end{equation}
For the attenuation of the link, we have used the same model proposed in~\cite{Ref18}
\begin{equation} \label{Lril}
\begin{split}
L\left ( r_{i,l} \right ) & = -20\log_{10}\left(\frac{4\pi}{\lambda} \right) \\
 & - 10n\left [ 1-b+\frac{bc}{\lambda f_{0}} \right ]\log_{10}\left ( r_{i,l} \right )-X_{\sigma } 
\end{split}
\end{equation}
where $n$ is the path loss exponent, $b$ is a system parameter, $f_0$ is the fixed reference frequency (the centroid of all frequencies represented by the path loss model). Finally, $X_{\sigma }$ is the shadow fading term with zero mean and variance $\sigma^2$. Parameters values of Eq.~\ref{Lril} for different indoor propagation scenarios have been defined in~\cite{Ref18}. \\
Millimeter wave frequencies are known for their quasi optical nature~\cite{Malt10}, and similar like light multipath components could be blocked. Blockage modeling of LOS component is of fundamental interest as it caries the highest energy. Therefore, channel model for LOS component is defined as  
\begin{equation}
\label{HLOSI}
\begin{array} {lcl}
\bm{H}_\text{LOS}\left(\tau \right) & = & I_\text{LOS}\left ( d \right )\sqrt{Kn_\text{BS}}e^{j\eta }\sqrt{L\left ( d \right )} \times \\ &  &  \bm{a}_r\left( \phi_{i,l}^A , \theta_{i,l}^A\right)\bm{a}_t^H\left( \phi_{i,l}^D,\theta_{i,l}^D\right) h\left( \tau -\tau_{i,l}\right)   
\end{array}
\end{equation}
where, $\eta\sim \mathcal{U}\left(0,2\pi \right) $ corresponds to the phase of the LOS component and  $I_\text{LOS}\left(d \right) $ is a random variable indicating if a LOS link is present between the transmitter and receiver at a distance $d$. Regarding the probability $ p $  of a LOS link, we used the same model defined in~\cite{Ref18,Ref19} for the evaluation of  
 considered  shopping mall scenario.
Some other selected statistical features and model distributions are summarized in Table.~\ref{statTab} whereby other detailed description are available in~\cite{BuzziD16}. \\
We have placed different number of users in a ring of radius $R$. Inside the ring, location of the users and their antenna orientations changes following a uniform random distribution in each channel snapshot. All users are assumed to be static for a particular channel realization, hence the channel stays time invariant. Multiuser MIMO channel matrix is now defined as 
\begin{equation}
\label{HLTI2}
\bm{H}=\left[  \bm{H}_1\quad \bm{H}_2\cdots\bm{H}_{n_U} \right] \in  \mathbb{C}^{n_\text{BS}\times n_\text{U}K} 
\end{equation}
\begin{table}[!t]
\centering
\caption{Table of statistical distributions and their parameters along with some other miscellaneous parameters}
\label{statTab}
\begin{tabular}{|l|l|l|}
\hline
\rowcolor[HTML]{00D2CB} 
Parameter & Distribution            & Mean \& St. dev.           \\ \hline
Azi. AOA per cluster $i$   & Laplacian& $ \theta_i^A \sim \mathcal{U}[0,2\pi] $ \&  $ 5^{\circ}  $  \\ \hline
Azi. AOD per cluster $i$   & Laplacian& $\theta_i^D\sim \mathcal{U}[\frac{-\pi}{2},\frac{\pi}{2}] $ \& $5^{\circ}$\\ \hline
Ele. AOA per cluster $i$   & Laplacian & $\phi_i^A\sim \mathcal{U} [\frac{-\pi}{2},\frac{\pi}{2}] $ \& $5^{\circ}$\\ \hline
Ele. AOD per cluster $i$  & Laplacian & $\phi_i^D\sim \mathcal{U} [\frac{-\pi}{2},\frac{\pi}{2}]$ \& $5^{\circ}$\\ \hline
\# Clusters $\left(  N_{cl}\right) $   & \multicolumn{2}{l|}{$ \max \left\lbrace Poisson(\Lambda),1\right\rbrace,\Lambda=1.9$\cite{Ref7RapaportmmCapacity} }           \\ \hline
\# Rays per cluster $i$  & \multicolumn{2}{l|}{$\mathcal{U} [1,30]$ \cite{Ref14GlobcomRapaport}}           \\ \hline
\multicolumn{3}{|l|}{\textbf{Miscellaneous parameters}}                        \\ \hline
\rowcolor[HTML]{00D2CB} 
Parameter & \multicolumn{2}{l|}{\cellcolor[HTML]{00D2CB}Value} \\ \hline
Scenario     & \multicolumn{2}{l|}{Indoor Shopping Mall}                     \\ \hline
Carrier frequency     & \multicolumn{2}{l|}{73 GHz}                     \\ \hline
$n_\text{BS}$& \multicolumn{2}{l|}{($5\times8 $,$20\times8 $) UPA $ \rightarrow$(40,160) ant }                     \\ \hline
$ n_\text{U} $   & \multicolumn{2}{l|}{2,5,10}                         \\ \hline
Antenna spacing $ d_\text{a} $   & \multicolumn{2}{l|}{$0.5 \lambda, 4 \lambda,6 \lambda$}                         \\ \hline
\# antennas per user $K$      & \multicolumn{2}{l|}{1}                              \\ \hline
$ h_\text{BS} $     & \multicolumn{2}{l|}{7 m }                     \\ \hline
$ h_\text{MU} $     & \multicolumn{2}{l|}{1.68 m }                     \\ \hline
Average BS-MU distance    & \multicolumn{2}{l|}{$20$ m}                          \\ \hline
Radius of the ring $\left( R\right) $ & \multicolumn{2}{l|}{$5$ m}                              \\ \hline
BS orientation    & \multicolumn{2}{l|}{Arbitrary}                              \\ \hline
MS orientation    & \multicolumn{2}{l|}{Arbitrary}                              \\ \hline

\end{tabular}
\end{table}

\section{Sum rate calculation with zero forcing precoder}
Considering equal power allocation per user, the zero-forcing  precoding matrix is defined as $\boldsymbol{W}=\boldsymbol{H}^+$, where $\boldsymbol{H}^+$ is the pseudo inverse of $\boldsymbol{H}$ and $\boldsymbol{w}_i \in \boldsymbol{W}$. The post processing SNR of the $i^{th}$ user is defined as
\begin{equation}
\label{OSNR}
   \text{SNR}^\text{ZF}_i = \frac{P}{N_0L_\text{U}\left[ \left( \boldsymbol{H}^H\boldsymbol{H}\right)^{-1}\right]_{i,i}  }
\end{equation}
Where $ L_\text{U} $ corresponds to the number of selected users. The overall sum rate $R^\text{ZF}$ of the selected users using zero-forcing beamforming is defined in~\cite{CS03} and it can be written as     
\begin{equation}
\label{SRate}
   R^\text{ZF} =\sum\limits_{i=1}^{K} \log_2\left(\text{SNR}^\text{ZF}_i \right) \cdotp
\end{equation}    
\section{Simulation results}
\label{sec:Simulation} 
\begin{figure}[t!]
\begin{center}
\includegraphics[width=0.95\columnwidth]{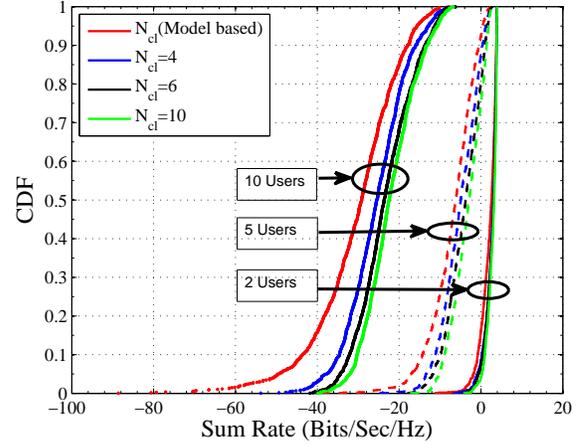}
\end{center}
\caption{Sum rate by changing the number of clusters (SNR= 0 dB, $ d_a=0.5\lambda$, $n_\text{BS}=20\times8 $ UPA)}
\label{ClustringNbs160DefspacingSNR0db}
\end{figure}

\begin{figure}[t!]
\begin{center}
\includegraphics[width=0.95\columnwidth]{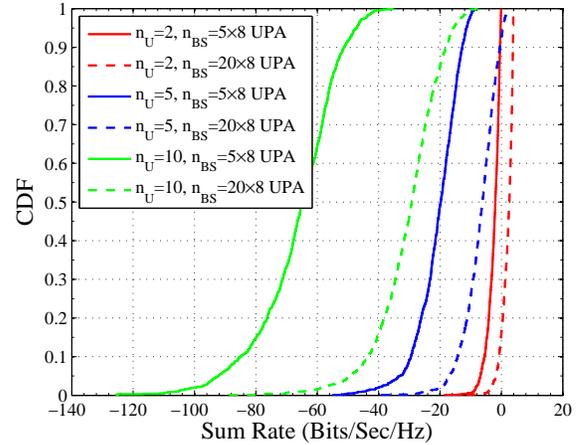}
\end{center}
\caption{Sum rate by changing the number of BS antennas $ N_{cl} -$ Model based, SNR=0 dB, $ d_a=0.5\lambda$}
\label{NbsDefclusterDefspacingSNR0db}
\end{figure}

\begin{figure}[t!]
\begin{center}
\includegraphics[width=0.95\columnwidth]{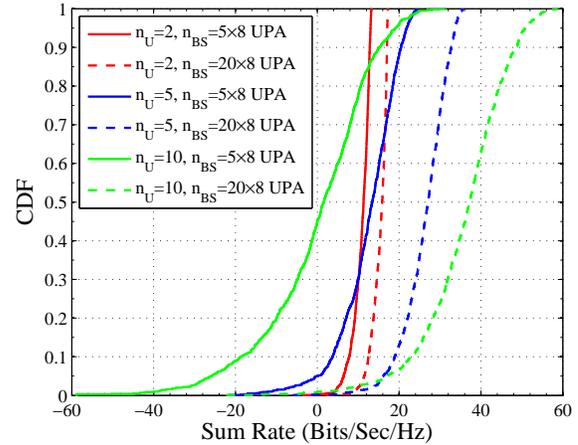}
\end{center}
\caption{Sum rate by changing the number of BS antennas $ N_{cl} -$ Model based, SNR=20 dB, $ d_a=0.5\lambda$}
\label{NbsDefclusterDefspacingSNR20db}
\end{figure}

\begin{figure}[t!]
\begin{center}
\includegraphics[width=0.95\columnwidth]{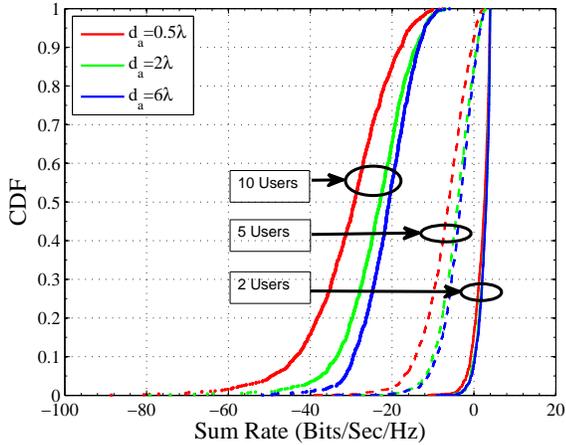}
\end{center}
\caption{Sum rate by changing the BS antenna spacing $ N_{cl} -$ Model based, SNR=0 dB, $ d_a=0.5\lambda$, $n_\text{BS}=20\times8 $ UPA}
\label{SpacingNbs160DefClustringSNR0db}
\end{figure}

\begin{figure}[t!]
\begin{center}
\includegraphics[width=0.95\columnwidth]{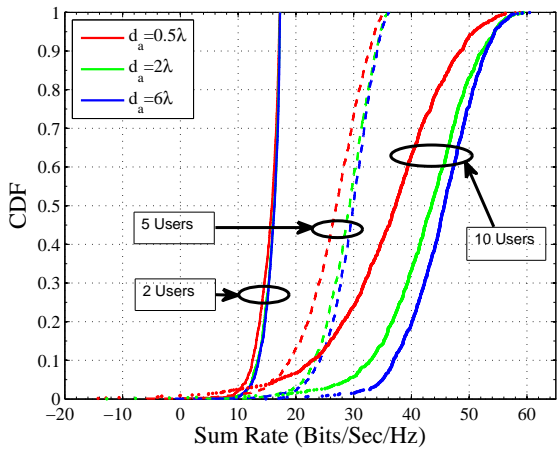}
\end{center}
\caption{Sum rate by changing the BS antenna spacing $ N_{cl} -$ Model based, SNR=20 dB, $ d_a=0.5\lambda$, $n_\text{BS}=20\times8 $ UPA}
\label{SpacingNbs160DefClustringSNR20db}
\end{figure}

In the following, we evaluate the impact of three channel parameters: number of clusters, number of BS antennas in the uniform patch array (UPA) and inter element distance (IED) of BS antennas on sum rate of $ n $ users at different SNR setups.\\
In Fig.~\ref{ClustringNbs160DefspacingSNR0db}, we compare the impact of multipath  clusters at low SNR i.e. at 0 dB. It can be seen that the sum rate increases by increasing the number of clusters for ten and five users but increase in sum rate is negligible for two users. It can also be observed that increase in the sum rate is more significant for 10 users. Fig.~\ref{ClustringNbs160DefspacingSNR0db} also shows that even if the number of clusters are increased, the sum rate of ten users setup is still lower than two users setup at low SNR regime. \\
\begin{figure}[t!]
\begin{center}
\includegraphics[width=0.95\columnwidth]{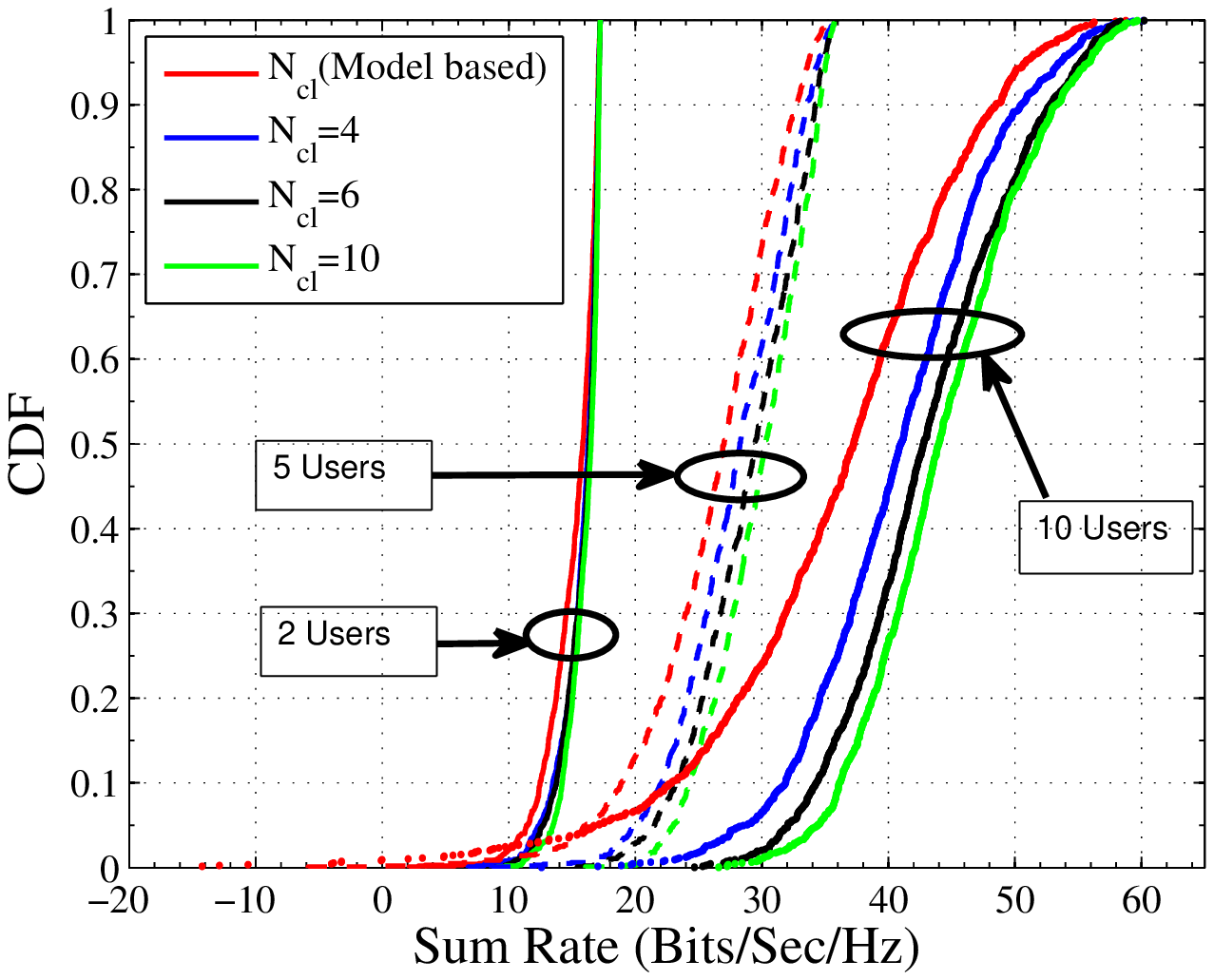}
\end{center}
\caption{Sum rate by changing the number of clusters (SNR= 20 dB, $ d_a=0.5\lambda$, $n_\text{BS}=20\times8 $ UPA)}
\label{ClustringNbs160DefspacingSNR20db}
\end{figure}
\begin{figure}[t!]
\begin{center}
\includegraphics[width=0.95\columnwidth]{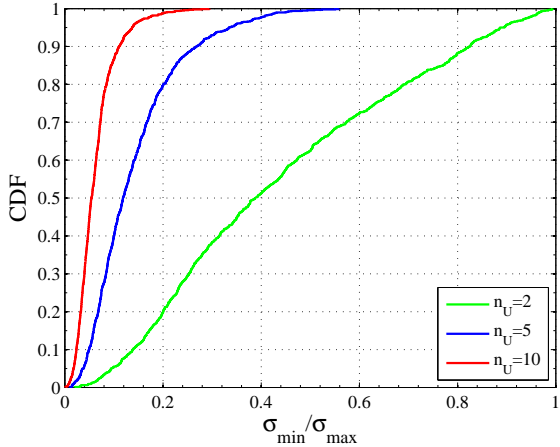}
\end{center}
\caption{Channel condition number by changing the number of users,  $ N_{cl}- $Model based, $ d_a=0.5\lambda$, $n_\text{BS}=20\times8 $ UPA}
\label{UsersClustDefLambDef}
\end{figure}
At high SNR dynamics i.e. at 20 dB, Fig.~\ref{ClustringNbs160DefspacingSNR20db} shows that the sum rate of ten users is significantly higher as compared to two and five users. This is due to higher multiplexing gains at high SNR regime. The analysis of channel condition number $ \frac{\sigma_{\min}}{\sigma_{\max}} $ from Fig.~\ref{UsersClustDefLambDef} shows that the CDF of two users is more closer to $'1'$ which shows singular value spread is lower for 2 users as compared to 5 and 10 users where it is high enough. It indicates that the channel correlation among users increases by increasing the number of users so the spatial separation among the users becomes increasingly difficult. Hence, the \textit{channel hardening} assumption is increasingly violated with increasing number of users in the ring. Channel hardening refers to the phenomenon where the off-diagonal terms of the ${\bm H}^H{\bm H}$ matrix become increasingly weaker compared to the diagonal terms as the size of the channel gain matrix ${\bm H}$ increases \cite{hardeningDef}. \\ 
It is generally assumed that with large number of antenna elements at the BS increases the channel hardening factor so that the low complexity match filtering precoder i.e. ${\bm W}=\bm{H}^H$ becomes optimal. Mathematically, it means that ${\bm W}{\bm H} $ is an identity matrix. However, high correlation in Fig.~\ref{UsersClustDefLambDef} shows that  ${\bm W}=\bm{H}^H$ may not result in identity matrix that makes match filtering sub-optimal. So, one can use other linear but relatively more computationally complex precoding schemes such as zero forcing (ZF) or minimum mean square error (MMSE) techniques in conjunction with multiuser selection.
User selection techniques  are well understood to be an effective way to reduce the computational complexity of the system. While optimal user selection is a NP-hard \cite{CivrilM09} problem and requires exhaustive search over all possible user combinations. Therefore, sub-optimal greedy incremental and decremental selection strategies are widely used.\\
The linear gain in channel capacity  by increasing the number of antennas depends upon the SNR scaling and the number of scattering points in the channel.  Increased number of scatterers give rise to the higher  number of multipath components. Therefore, leading towards multipath diversity which in turn results in increased channel capacity. In general, at lower frequencies ($ < 6 $ GHz),  the famous 3GPP channel models e.g. WINNER/SCM \cite{WINNER-II} assume that resolvable multipath components are composed of 20 sub-paths so that central limit theorem holds and the amplitude fading of the resolvable paths follows the Rayleigh distribution. If the number of multipath components are such high then off-course the high spatial resolution associated with large number of antenna elements in the array may explore more scattering points in the channel and hence may increase the channel capacity. However, on the other hand mmWave channels could be sparse due to higher propagation and penetration losses from the different objects in the channel. Therefore, it is quite likely that under certain scenarios the number of antenna elements in the MIMO array is larger than the number of resolvable multipath components. In such scenarios, further increase in the antenna elements may not result in  linear gain in the channel capacity and in the worst case it could be bottlenecked.  
Fig.~\ref{NbsDefclusterDefspacingSNR0db} shows the CDF analysis of the sum rate  by changing the number of BS antennas from $ 5\times 8 $ UPA (40 antennas) to $ 20\times 8 $ UPA (160 antennas) for two, five and ten users at 0 dB SNR. The sum rate of two users is higher than five and ten users because the multiplexing gain is negligible at 0 dB. However, the change in sum rate is significant for ten users as compared to five and two users by increasing the BS antennas due to exploration of more multipath for more users. The results from Fig.~\ref{NbsDefclusterDefspacingSNR20db} at 20 dB SNR shows that the sum rate of ten users is higher than five and two users. At high SNR, increasing BS antennas gives rise to multiplexing gain.\\
The separation between the antenna elements explains the channel correlation properties and the exploration of the number of  scatterers. When antenna elements are very close to each other, the channel correlation will be very high and the interference among the users will increase. From Fig.~\ref{SpacingNbs160DefClustringSNR0db}, we can see that increasing inter element distance (IED) at BS results more change in sum rate for ten users because of reduction in channel correlation. However, from Fig.~\ref{SpacingNbs160DefClustringSNR0db} at 0 dB SNR, one can see that the even by increasing the IED the sum rate of ten users is significantly lower than five and two users because the diversity gain is high in case of two and five users as compared to ten users. 
From Fig.~\ref{SpacingNbs160DefClustringSNR20db} at 20 dB SNR, by changing $ d_a=0.5\lambda $ to $ d_a=6\lambda $  gives minor gain in sum rate foe two and five users setups. Whereas, the weaker eigen mode are scaled up by high SNR so the sum rate of ten users significantly  increase by increasing the IED due to multiplexing gain. The gain in sum rate  by increasing the IED is limited up to the extent where the number of scatterers can be fully exploited and further increase in IED of BS antennas could not provide the gain in sum rate of mmWave multiuser system.


\section{Conclusion and future work}
\label{sec:CONCLUSION}
In this paper, we have conducted several experiments, where we have increased the number of multipath clusters in the channel and additionally at the base station antenna array we increased inter-antenna element distance and the number of antenna elements. Our experiments used the statistical channel models derived from the realistic state of the art mmwave channel models. Results show that the mmwave frequencies channels are correlated in space and it is hard to spatially separate users in the dense cellular networks. In contrast to centimeter wavelength frequencies, the results of our experiments show that the channel hardening assumption does not remain valid at the mmwave channels. We have provided evidence in terms of channel conditions and sum rate performance. Hence, for the sum rate maximization we recommend to use multiusers selection and scheduling algorithms.    

\bibliographystyle{IEEEtran}


\end{document}